\documentclass[11pt]{article}
\usepackage{moriond,epsfig}

\def\Journal#1#2#3#4{{#1} {\bf #2}, #3 (#4)}

\def\NPB{{\em Nucl. Phys.} B}
\def\PLB{{\em Phys. Lett.}  B}

\def\PRD{{\em Phys. Rev.} D}
\def\ZPC{{\em Z. Phys.} C}
\def\PR{\em Phys. Rept.}
\def\JHEP{\em JHEP}
\def\EPJC{{\em Eur. Phys. J.} C}

\def\be{\begin{equation}}
\def\ee{\end{equation}}
\def\bea{\begin{eqnarray}}
\def\eea{\end{eqnarray}}

\begin{document}
\vspace*{4cm}
\title{SU(5) \& A(4)}

\author{ ALFREDO URBANO }

\address{Dipartimento di Fisica, Universit\`{a} del Salento \& INFN, sez. di Lecce\\
via per Arnesano I, 73100 Lecce, Italia}

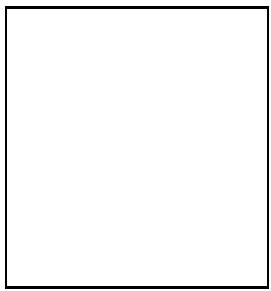
\maketitle\abstracts{
The introduction of a Flavour Symmetry can represent an interesting way in which one can try to find an answer to some intriguing problems in Flavour Physics, like the hierarchy between the fermion masses or the particular values of mixing angles. In the meantime the necessity to set this symmetry in a realistic context grows up; this context should be able to enlarge our incomplete knowledge of fundamental interactions as described in the framework of the Standard Model. Following this direction a merging between $\mbox{A}_{4}$ and $\mbox{SU}(5)$ can be possible.\\
}

\section{Introduction}

The $\mbox{SU}(5)$ gauge group represents the minimal playground in which one can try to enlarge the $\mbox{SU}(3)_{C}\otimes\mbox{SU}(2)_{L}\otimes\mbox{U}(1)_{Y}$ symmetry group of the Standard Model (SM), searching for an high energy unified description of strong and electroweak fundamental interactions. The correct realization of this picture, as suggested by the apparent convergence of SM couplings, could lead to important consequences both from a theoretical as well as a phenomenological point of view as, e.g., the explanation of electric charge quantization or the prediction of the proton decay. On the other side the existence of neutrino masses and oscillations, the experimental bounds on proton lifetime and the necessity to obtain a phenomenologically valid description for masses and mixings are able to rule out a large number of models. Presently in literature there are just two viable models, in the context of $\mbox{SU}(5)$, that can take into account proprieties of renormalizability as well as a realistic description of high energy phenomenology in a ``minimal" way; in the first one \cite{Perez:2007iw}, called \emph{Supersymmetric Adjoint SU(5)} ($\mathcal{SA}$), it's possible to obtain a consistent and predictive description of neutrino masses through the presence of both the Type I \cite{Albright:2004kb} and the Type III \cite{Foot:1988aq} SeeSaw. In the second one \cite{Dorsner:2007fy}, instead, the neutrino phenomenology is obtained through the implementation of the so called Type II \cite{Mohapatra:1999zr} SeeSaw mechanism.\\
In the following I take into account both these models trying to analyze the possible role that could play the introduction of a Flavour Symmetry (FS) represented by the discrete group of the regular tetrahedron, $\mbox{A}_4$.
In section \ref{sec:gen} I take into account briefly the main proprieties of the two renormalizable supersymmetric $\mbox{SU}(5)$ framework that I consider as background for the analysis accomplished in section \ref{sec:fla} with the introduction of the $\mbox{A}_4$ group as an exact flavour symmetry at GUT scale. Section \ref{sec:concl} is left to conclusions.

\section{Renormalizable SUSY $\mbox{SU}(5)$ GUT models}\label{sec:gen}

Following ref. \cite{Dimopoulos:1981zb,Slansky:1981yr} it's well known that the chiral superfields of the Minimal Supersymmetric SM (MSSM) find their correct embedding inside the $\mbox{SU}(5)$ structure considering four different irreducible representations: $\overline{\textbf{5}}_{\tiny\textbf{T}} =(D^{C},L)$, $\textbf{10}_{\tiny\textbf{T}}=(U^{C},Q,E^{C})$, $\overline{\textbf{5}}_{\tiny\textbf{H}}=(T_{\overline{5}},H_{\overline{5}})$ and $\textbf{5}_{\tiny\textbf{H}}=(T_{5},H_{5})$,
where the tiny subindex $_{\tiny{\textbf{T}}}$ refers to matter (for a single family) while $_{\tiny{\textbf{H}}}$ to Higgs chiral superfields, and where I have identified in the left hand side the superfield name with the corresponding $\mbox{SU}(5)$ representation number.
In addiction we need also an extra Higgs chiral superfield in the adjoint representation $\textbf{24}_{\tiny{\textbf{H}}}$ in order to break $\mbox{SU}(5)$ in the SM gauge group.
Unfortunately this beautiful and minimal unification picture is destroyed by our knowledge of the low energy phenomenology because there's no renormalizable way to obtain a neutrino mass matrix $\mbox{M}_{\nu}$ and because with this field content a wrong relation between charged lepton and down-type quark mass matrices, i.e. $\mbox{M}^{T}_{E}=\mbox{M}_{D}$, appears at GUT scale. In order to fix this last problem keeping untouched the renormalizability of the theory, it's possible to consider in the Higgs sector \cite{Georgi:1979df} two extra chiral superfields transforming under $\mbox{SU}(5)$ according to the $\textbf{45}_{\tiny{\textbf{H}}}$ and the $\overline{\textbf{45}}_{\tiny{\textbf{H}}}$ representations. The relevant superpotential for the Yukawa couplings is, introducing $i,j$ as flavour indices and considering R-parity: $
\mathcal{W}\ni \textbf{10}_{\tiny{\textbf{T}},i}\,\mbox{Y}_{\overline{5},ij}\,\overline{\textbf{5}}_{\tiny{\textbf{T}},j}\,
\overline{\textbf{5}}_{\tiny{\textbf{H}}}+
\textbf{10}_{\tiny{\textbf{T}},i}\,\mbox{Y}_{\overline{45},ij}
\,\overline{\textbf{5}}_{\tiny{\textbf{T}},j}\,
\overline{\textbf{45}}_{\tiny{\textbf{H}}}+
\textbf{10}_{\tiny{\textbf{T}},i}\,\mbox{Y}_{5,ij}\,\textbf{10}_{\tiny{\textbf{T}},j}\,
\textbf{5}_{\tiny{\textbf{H}}}+
\textbf{10}_{\tiny{\textbf{T}},i}\,\mbox{Y}_{45,ij}\,\textbf{10}_{\tiny{\textbf{T}},j}\,
\textbf{45}_{\tiny{\textbf{H}}}$.
After electroweak symmetry breaking  we have the following relations: $\mbox{M}_{E}=\mathcal{Y}_{\overline{5}}^{T}-6\mathcal{Y}_{\overline{45}}^{T}$\,, $\mbox{M}_{D}=\mathcal{Y}_{\overline{5}}+2\mathcal{Y}_{\overline{45}}$\, and
$\mbox{M}_{U}=4(\mathcal{Y}_{5}^{T}+\mathcal{Y}_{5})-8(\mathcal{Y}_{45}-\mathcal{Y}_{45}^{T})$, that allow a correct description of fermion masses (here, e.g., $\mathcal{Y}_{5}=\mbox{Y}_{5}\langle \textbf{5}_{\tiny{\textbf{H}}}\rangle$).
Until this point, however, neutrinos are still massless.
In order to realize the Type I and the Type III SeeSaw, there must be both a fermionic singlet as well as a fermionic triplet. Inside the context of $\mbox{SU}(5)$ it's natural \cite{Perez:2007iw} to introduce an extra chiral superfield of matter in the adjoint representation $\textbf{24}_{\tiny{\textbf{T}}}$ that contains both; the new terms in the superpotential are: $\mathcal{W}\ni \overline{\textbf{5}}_{\tiny{\textbf{T}},i}\,\mbox{Y}_{\nu,i}^{45}\,\textbf{24}_{\tiny{\textbf{T}}}\,\textbf{45}_{\tiny{\textbf{H}}}+
\overline{\textbf{5}}_{\tiny{\textbf{T}},i}\,\mbox{Y}_{\nu,i}^{5}\,\textbf{24}_{\tiny{\textbf{T}}}\,\textbf{5}_{\tiny{\textbf{H}}}+\mbox{M}_{24}\,
\textbf{24}_{\tiny{\textbf{T}}}\,\textbf{24}_{\tiny{\textbf{T}}}+\lambda_{24}\,\textbf{24}_{\tiny{\textbf{T}}}\,\textbf{24}_{\tiny{\textbf{T}}}\,
\textbf{24}_{\tiny{\textbf{H}}}
$. The neutrino mass matrix is obtained integrating out the heavy singlet $\rho_{0}$ and the neutral component of the triplet $\rho_{3}$ in $\textbf{24}_{\tiny{\textbf{T}}}$;
in this way it's possible to obtain both the normal hierarchy as well as the inverted one, predicting also one massless neutrino.
In the context of the Type II SeeSaw, instead, the neutrinos acquires a mass through the coupling of the leptonic doublet $L$ with an heavy scalar triplet $T$. In a $\mbox{SU}(5)$ framework  it's natural \cite{Dorsner:2007fy} to introduce two extra chiral superfield in the Higgs sector transforming according to $\textbf{15}_{\tiny{\textbf{H}}}$ (that contains $T$) and $\overline{\textbf{15}}_{\tiny{\textbf{H}}}$  representations.
The extra terms in the superpotential of the theory are: $\mathcal{W}\ni \overline{\textbf{5}}_{\tiny{\textbf{T}},i}\,Y^{15}_{\nu,ij}\,\textbf{15}_{\tiny{\textbf{H}}}\,\overline{\textbf{5}}_{\tiny{\textbf{T}},j}+\mbox{M}_{15}\,
\textbf{15}_{\tiny{\textbf{H}}}\,\overline{\textbf{15}}_{\tiny{\textbf{H}}}+h_{24}\,
\textbf{15}_{\tiny{\textbf{H}}}\,\textbf{24}_{\tiny{\textbf{H}}}\,\overline{\textbf{15}}_{\tiny{\textbf{H}}}+
h_{5}\,\overline{\textbf{15}}_{\tiny{\textbf{H}}}\,\textbf{5}_{\tiny{\textbf{H}}}\,\textbf{5}_{\tiny{\textbf{H}}}+
h_{\overline{5}}\,\textbf{15}_{\tiny{\textbf{H}}}\,\overline{\textbf{5}}_{\tiny{\textbf{H}}}\,\overline{\textbf{5}}_{\tiny{\textbf{H}}}+
h_{45}\,\overline{\textbf{15}}_{\tiny{\textbf{H}}}\,\textbf{45}_{\tiny{\textbf{H}}}\,\textbf{45}_{\tiny{\textbf{H}}}+
h_{\overline{45}}\,\textbf{15}_{\tiny{\textbf{H}}}\,\overline{\textbf{45}}_{\tiny{\textbf{H}}}\,\overline{\textbf{45}}_{\tiny{\textbf{H}}}$; integrating out the heavy scalar triplet it's possible to obtain a neutrino Majorana mass matrix $M_{\nu,ij}=\frac{\Lambda^{2}}{\mbox{M}_{15}}Y^{15}_{\nu,ij}$, with $\Lambda$ of the order of the electroweak scale.

\section{The $\mbox{SU}(5)\otimes \mbox{A}_{4}$ Renormalizable SUSY GUT theory}\label{sec:fla}



$\mbox{A}_{4}$ is the group of even permutations of $4$ objects and it's also isomorphic to the tetrahedral group;
for a deeper understanding of $\mbox{A}_{4}$ as a FS group I redirect the interested reader to the milestone \cite{Babu:2002dz}. Here I remember just that $\mbox{A}_{4}$ has three non equivalent one dimensional representations $\texttt{1}$, $\texttt{1}^{\prime}$ and $\texttt{1}^{\prime\prime}$ and a three dimensional one $\texttt{3}$.
As a FS, $\mbox{A}_{4}$ have to be imposed as a symmetry of the entire superpotential $\mathcal{W}$ at GUT scale. \\
In \cite{Ciafaloni:2009ub} we have tried to set the $\mbox{A}_{4}$ FS in the $\mathcal{SA}$ context; the only viable situation is the one in which all the chiral superfield of the theory are $\mbox{A}_{4}$ triplets except for $\textbf{24}_{\tiny{\textbf{H}}}$
that is a singlet
\footnote{There's also another realistic $\mbox{A}_{4}$ scenario, that is the one in which $\overline{\textbf{5}}_{\tiny{\textbf{T}}}^{a=1,2,3}\sim \texttt{1},\texttt{1}',\texttt{1}''$; this situation is considered also in \cite{Altarelli:2008bg} even if into a $5$-dim background.}. Moreover the most important point concerns the alignments of the Vacuum Expectation Values (VEVs) in the flavour space; it's possible to verify that, in order to have the possibility to fit correctly masses and mixings, the only choice is \footnote{Obviously in order to have a realistic situation it's necessary to show that the considered alignments are in agreement with the minimization of the scalar potential.}: $\langle \overline{\textbf{5}}_{\tiny{\textbf{H}}}\rangle \propto \langle \textbf{5}_{\tiny{\textbf{H}}}\rangle=v_{5}(1,0,0)^{T}$,
$\langle \overline{\textbf{45}}_{\tiny{\textbf{H}}}\rangle=v_{\overline{45}}(1,1,1)^{T}$ and $\langle\textbf{45}_{\tiny{\textbf{H}}}\rangle=(v_{45},\delta v_{45},\delta v_{45})^{T}$;
with these assumptions, in fact, we are able to fit the eigenvalues of
the mass matrices for $E=(e,\mu,\tau)$, $D=(d,s,b)$ and $U=(u,c,t)$
with the running quark and lepton masses of \cite{Das:2000uk}. The fit is acceptable ($\chi^{2}/n.dof\sim 1.5$) and it's also possible at this stage to estimate numerically the Cabibbo-Kobayashi-Maskawa (CKM) matrix, in good agreement with its perturbative structure.
In the leptonic sector the situation of mixing is more complicated. Considering $\textbf{24}_{\tiny{\textbf{T}}}$ as an $\mbox{A}_{4}$ triplet, in fact, it's possible to have a structure dictated by the presence of the FS just for those Dirac matrices that follow directly from the Lagrangian, namely $\mbox{Y}_{\nu}^{45}$ and $\mbox{Y}_{\nu}^{5}$ in the interactions: $\overline{\textbf{5}}_{\tiny{\textbf{T}}}\,\mbox{Y}_{\nu}^{45}\,\textbf{24}_{\tiny{\textbf{T}}}\,\textbf{45}_{\tiny{\textbf{H}}}+
\overline{\textbf{5}}_{\tiny{\textbf{T}}}\,\mbox{Y}_{\nu}^{5}\,\textbf{24}_{\tiny{\textbf{T}}}\,\textbf{5}_{\tiny{\textbf{H}}}$; this structure, however, gets completely lost performing the SeeSaw mechanism.  Roughly speaking, in fact, it requires the combination $\mbox{M}_{Dirac}\mbox{M}_{Majorana}^{-1}\mbox{M}_{Dirac}^{T}$ for both the Type I as well as the Type III case,
leaving us with an almost generical Majorana mass matrix $\mbox{M}_{\nu}$ which possesses however the remarkable property to fit the tri-bimaximal (TBM) hypothesis \cite{Harrison:2002er} (see eq. \ref{eq:tbmhypothesis}). \\
Considering the $\mbox{SU}(5)$ framework with the Type II SeeSaw, instead,
 both chiral matter supermultiplets $\overline{\textbf{5}}_{\tiny{\textbf{T}},i}$ and $\textbf{10}_{\tiny{\textbf{T}},i}$ transform under the $\mbox{A}_{4}$ group according to its three dimensional representation, obtaining again an original approach if compared with the one in \cite{Altarelli:2008bg}; with this assumption, therefore, the only way in which to obtain a realistic description requires the presence of reducible four dimensional representations of $\mbox{A}_{4}$. Considering the usual irreducible ones, in fact, it's not possible to fit correctly the quark and lepton running masses. Specifically we must have: $\overline{\textbf{5}}_{\tiny{\textbf{H}}},\textbf{5}_{\tiny{\textbf{H}}},
\overline{\textbf{45}}_{\tiny{\textbf{H}}},\textbf{45}_{\tiny{\textbf{H}}}\sim \texttt{4}$, $\overline{\textbf{15}}_{\tiny{\textbf{H}}}\sim\texttt{4}''$ and $\textbf{15}_{\tiny{\textbf{H}}}\sim\texttt{4}'$. Obviously such an Higgs sector might sound cumbersome; here I want to stress that it's the only way in which try to preserve the beautiful features of $\mbox{A}_{4}$ group as a FS without losing the renormalizability of the entire theory in a $4$-dim background. Another point to discuss concerns the minimization of the scalar potential; leaving a deeper analysis to a forthcoming paper \cite{Ciafaloni:2009prox}, it's possible to prove that under certain assumptions the VEVs of the triplets in $\overline{\textbf{5}}_{\tiny{\textbf{H}}}$ and $\overline{\textbf{45}}_{\tiny{\textbf{H}}}$ as well as the ones in $\textbf{5}_{\tiny{\textbf{H}}}$ and $\textbf{45}_{\tiny{\textbf{H}}}$ have to be aligned among each other in the flavour space, with the natural choice: $\langle \textbf{5}_{\tiny{\textbf{H}}}\rangle_{\tiny{\texttt{3}}}=v_{5}(1,1,1)^{T}$ (analogous relations holds for the other triplets; here the subindex $_{\tiny{\texttt{3}}}$ refers to the triplet in $\texttt{4}=\texttt{3}\oplus\texttt{1}$).
As a consequence the quark and lepton mass matrices acquire the form:
\begin{equation}\label{eq:matricef}
\mbox{M}_{F}=\left(
               \begin{array}{ccc}
                 h_{F}^{0} & A_F & B_F \\
                 B_F & h_{F}^{0} & A_F \\
                 A_F & B_F & h_{F}^{0} \\
               \end{array}
             \right)=\mbox{U}_{\omega}\mbox{M}_{F}^{\Delta}\mbox{U}_{\omega}^{\dag},\,\,\,\,\mbox{with:}\,\,\mbox{U}_{\omega}=\frac{1}{\sqrt{3}}
             \left(
               \begin{array}{ccc}
                 \omega & \omega^{2} & 1 \\
                 \omega^{2} & \omega & 1 \\
                 1 & 1 & 1 \\
               \end{array}
             \right),\,\,\omega^{3}=1,
\end{equation}
where $F=E,D,U$ and where $h_{F}^{0}$, $A_{F}$ and $B_{F}$ are functions of the VEVs and of the parameters in the superpotential, independent among each other for different values of $F$. The matrices in (\ref{eq:matricef}) cannot explain the hierarchy but allow a fit of the running lepton and quark masses for each value of $F$, as shown in \cite{Morisi:2007ft}. At this step the CKM matrix is just the identity but it's straightforward to generate the Cabibbo's angle considering small perturbations in the vacuum alignments.
The mixing in the leptonic sector depends on the neutrino mass matrix. Supposing that $\langle\textbf{15}_{\tiny{\textbf{H}}}\rangle_{\tiny{\texttt{3}}}=v_{15}(0,0,1)^{T}$ it's possible to verify that the Majorana mass matrix has the form:
\begin{equation}\label{eq:majo}
\mbox{M}_{\nu}=\left(
                 \begin{array}{ccc}
                   a & b & 0 \\
                   b & a\omega & 0 \\
                   0 & 0 & a\omega^{2} \\
                 \end{array}
               \right)=\mbox{V}^{*}\mbox{M}_{\nu}^{\Delta}\mbox{V}^{\dag},\,\,\,\,\mbox{with:}\,\,\mbox{V}=\left(
                                                                                                             \begin{array}{ccc}
                                                                                                               \frac{\omega}{\sqrt{2}} & 0 & -\frac{i\omega}{\sqrt{2}} \\
                                                                                                               \frac{\omega}{\sqrt{2}} & 0 & \frac{i\omega}{\sqrt{2}} \\
                                                                                                               0 & 1 & 0 \\
                                                                                                             \end{array}
                                                                                                           \right);
\end{equation}
the mixing matrix for leptons reproduces exactly the TBM hypothesis:
\begin{equation}\label{eq:tbmhypothesis}
\mbox{U}^{\dag}_{\omega}\mbox{V}=\left(
                                                      \begin{array}{ccc}
                                                         \frac{2}{\sqrt{6}} & \frac{1}{\sqrt{3}} & 0 \\
                                                        -\frac{1}{\sqrt{6}} & \frac{1}{\sqrt{3}} & -\frac{1}{\sqrt{2}} \\
                                                         -\frac{1}{\sqrt{6}} & \frac{1}{\sqrt{3}} & \frac{1}{\sqrt{2}} \\
                                                       \end{array}
                                                     \right)\equiv\mbox{U}_{TBM}.
\end{equation}
The eigenvalues of $\mbox{M}_{\nu}$ in (\ref{eq:majo}) are \cite{Altarelli:2005yp}: $m_{1}=c+b,m_{2}=c,m_{3}=-c+b$, with $c\equiv\omega^{2}a $. The model predicts a normal hierarchy  and, if $\cos\theta_{cb}=-1$: $|m_{3}|\approx 0.053$ eV, $|m_{1}|\approx |m_{2}|\approx 0.017$ eV.

\section{Conclusions}\label{sec:concl}

 We have analyzed the possibility to consider the role of the $\mbox{A}_{4}$ FS inside a SUSY GUT based on $\mbox{SU}(5)$ gauge group. Contrarily to how done in \cite{Altarelli:2008bg} we have considered as guideline for this discussion the necessity to preserve the renormalizability of the entire model inside a $4$-dim space-time framework, taking into account two different possibilities in order to obtain the neutrino masses: in the first case an extra chiral supermultiplet in the matter sector allows to have an hybrid Type I\,+\,Type III SeeSaw while in the second case an extra chiral supermultiplet in the Higgs sector allows the Type II SeeSaw. Conclusions are very different. In the first scenario the masses and the mixings are obtained through a fit of the parameters, losing in this way the main \emph{raison d'\^{e}tre} of the $\mbox{A}_{4}$ FS but obtaining an almost degenerate spectrum for neutrinos with an absolute scale mass closer to the present experimental bounds \cite{Fogli:2008cx}. In the second scenario, however, the situation is completely different and the TBM mixing arises clearly.

\section*{Acknowledgments}
I'm indebted to Paolo Ciafaloni, Marco Picariello and Emilio Torrente-Lujan for the fruitful and precious collaboration. I warmly thank the Organizing Committee of the Moriond Conference for the welcoming and pleasant atmosphere breathed during the period spent in La Thuile.

\end{document}